# Evolutionary Optimization of a Charge Transfer Ionic Potential Model for Ta/Ta-Oxide Hetero-interfaces

Kiran Sasikumar,[1,ξ,*] Badri Narayanan,[1,ξ] Mathew Cherukara,[2] Alper Kinaci,[1] Fatih G. Sen,[1] Stephen K. Gray,[1,3] Maria K. Y. Chan,[1,3] and Subramanian K. R. S. Sankaranarayanan[1,3,*]

[1] Center for Nanoscale Materials, Argonne National Laboratory, Argonne, IL, 60439

[2] X-ray Sciences Division, Argonne National Laboratory, Argonne, IL, 60439

[3] Computation Institute, University of Chicago

[ξ] Equal Contributions

**Abstract**

Heterostructures of tantalum and its oxide are of tremendous technological interest for a myriad of technological applications, including electronics, thermal management, catalysis and biochemistry. In particular, local oxygen stoichiometry variation in $TaO_x$ memristors comprising of thermodynamically stable metallic (Ta) and insulating oxide ($Ta_2O_5$) have been shown to result in fast switching on the sub-nanosecond timescale over a billion cycles. This rapid switching opens up the potential for advanced functional platforms such as stateful logic operations and neuromorphic computation. Despite its broad importance, an atomistic scale understanding of oxygen stoichiometry variation across $Ta/TaO_x$ hetero-interfaces, such as during early stages of oxidation and oxide growth, is not well understood. This is mainly due to the lack of a unified interatomic potential model for tantalum oxides that can accurately describe metallic (Ta), ionic ($TaO_x$) as well as mixed ($Ta/TaO_x$ interfaces) bonding environments simultaneously. To address this challenge, we introduce a Charge Transfer Ionic Potential (CTIP) model for Ta/Ta-oxide system by training against lattice parameters, cohesive energies, equations of state (EOS), elastic properties, and surface energies of the various experimentally observed $Ta_2O_5$ polymorphs (hexagonal, orthorhombic and monoclinic) obtained from density functional theory (DFT) calculations. The best CTIP parameters are determined by employing a global optimization scheme driven by genetic algorithms followed by local Simplex optimization. Our newly developed CTIP potential accurately predicts structure, thermodynamics, energetic ordering of polymorphs, as well as elastic and surface properties of both Ta and $Ta_2O_5$, in excellent agreement with DFT calculations and experiments. We employ our newly parameterized CTIP potential to investigate the early stages of oxidation and atomic scale mechanisms associated with oxide growth on Ta surface at various temperatures. The CTIP potential developed in this work is an invaluable tool to investigate atomic-scale mechanisms and transport phenomena underlying the response of $Ta/TaO_x$ interfaces to external stimuli (e.g, temperature, pressure, strain, electric field etc.), as well as other interesting dynamical phenomena including the physics of switching dynamics in $TaO_x$ based memristors and neuromorphic devices.

---

[*] Corresponding authors: ssankaranarayanan@anl.gov, ksasikumar@anl.gov

**1. Introduction**

Tantalum and its oxide are materials of broad technological interest, owing to their exceptional physical, chemical, and opto-electronic properties [1, 2, 3, 4]. In the metallic form, tantalum and its alloys display excellent corrosion resistance [5], chemical inertness [6], high thermal stability [7], and exceptional strength [8]. This makes Ta alloys suitable for use in turbine blades, rocket nozzles and nose caps for supersonic aircraft [8]. As oxide, tantalum pentoxide (most stable $TaO_x$), is an important dielectric material that has widespread applications in capacitors, dynamic random access memories (DRAMs), optical coatings, high-temperature reflectors and antireflection coatings [4, 9, 10, 11, 12, 13]. Amongst the various dielectrics, $Ta_2O_5$ has received distinct attention owing to its high dielectric constant (~ 35), high refractive index and chemical and thermal stability, as well as the promise of compatibility with microelectronics processing [4, 9, 12, 14]. To successfully employ thin film materials such as $TaO_x$ for device applications, it is crucial to develop sophisticated synthesis and processing techniques, which in turn depends on our understanding of the structure-property relationships.

Hetero-structures of Ta and $Ta_2O_5$ are also being considered as potential platforms for neuromorphic computation [4, 15]. $TaO_x$ memristors have been shown to display unique switching properties on the sub-nanosecond timescale over more than a billion cycles [15]. Such reversibility was attributed to the motion of oxygen vacancies. The subsequent localized variations in oxygen stoichiometries in TaO$_x$ system, under an applied electric field, have been exploited to switch between the two thermodynamically stable states *i.e.* metallic Ta and insulating Ta$_2$O$_5$ [4, 15]. In order to realize the use of TaO$_x$ in emerging applications such as neuromorphic computing, it is highly desirable to identify optimal material systems (and their state variables), and develop predictive models of the underlying atomistic mechanisms governing their resistive switching. An atomistic scale understanding of the structure and dynamics across Ta/TaO$_x$ interfaces will allow us to predictively model and control the resistive state, thereby facilitating their integration into functional systems.

In addition to the metal/oxide interfaces and hetero-structures of TaO$_x$, the dynamical phenomena associated with the nanoscale oxidation [16, 17] and oxide growth processes in such systems are also of fundamental interest. Numerous experimental studies have focused on the thermal oxidation in bulk Ta and thin films of Ta [18, 19]. For example, Chandrasekaran *et al.* have used Auger depth profiling to study the thermal oxidation of 700 nm Ta thin films in the 600-1000 K range [18]. They find the oxide growth rate to be logarithmic at low temperatures <600 K and parabolic at high temperatures ~800 K. Similarly, Ruffel *et al.* have studied the formation and characterization of Ta$_2$O$_5$/TaO$_x$ (oxide/suboxide) hetero-structures via high fluence O ion-implantation into deposited Ta films [19]. Their study showed that oxygen stoichiometry in the hetero-structures can be tuned by O-ion implantation energy and fluence. While these studies have studied the microstructural evolution and formation of these oxide films, the nanoscale oxidation kinetics as well as atomistic details of the structure, stoichiometry and the limiting thickness of the Ta oxide thin films are largely unknown.

Probing the early stages of nanoscale oxidation and oxide growth is often difficult with experiments. With the significant improvement in computational resources, atomistic simulations are now emerging as a viable alternative to study the structural and dynamical evolution of the metal/oxide interfaces. First principles approaches based on density functional theories have been primarily employed to study the initial stages of O$_2$ reaction with metal surfaces, the dissociation processes involved in oxidation, and the stability of the various adsorption sites (top, bridge and hollow sites on FCC or HCP lattice) [20, 21]. Due to the computational cost, however, it is currently not feasible to use first principles approaches to directly model nanoscale oxide growth on metal surfaces. Classical molecular dynamics (MD) simulations based on semi-empirical potentials provide an ideal route to model dynamical phenomena associated with metal oxidation and oxide growth. There is, thus, a clear need for large-scale atomistic simulations employing semi-empirical potentials to study nanoscale oxidation of metals such as Ta.

The success of classical MD in modeling metal/oxide systems hinges on the ability of the employed potential model to accurately describe interatomic interactions in both the metallic and oxide regions. Additionally, this potential model should accurately capture the structural, chemical, thermodynamic, elastic, and surface properties of Ta and $TaO_x$ phases. To date, however, there is only one semi-empirical Morse-BKS potential function for Tantalum oxide, developed by Trinastic *et al*. [22], which primarily targets amorphous, yet stoichiometric,) $Ta_2O_5$ systems. This model is, however, a fixed charge model and is not capable of capturing the multiple oxidation states encountered in oxide/metal hetero-interfaces.

Here, we introduce a charge transfer ionic potential model for Ta/$TaO_x$ system that successfully captures the thermo-physical, structural and surface properties of both Ta and various polymorphs of $Ta_2O_5$ as well as their interfaces. For the potential formalism, we choose the charge transfer ionic interatomic potential developed by Zhou and Wadley [23, 24] since it allows the environment-dependent charges on the atoms to be dynamically deduced. During metal oxidation and oxide growth, metal ions attain significant positive charges whereas significant negative charges are attained by oxygen anions. Therefore, the atomic charges are environment dependent and dynamically vary during the course of the oxidation process. For instance, Ta charges in the stoichiometric oxide are expected to change continuously from a zero value in a fully metallic region to their valency-determined maximum value. To model such a scenario, one requires a potential model such as charge transfer ionic potential (CTIP) that seamlessly allows for switching between an environment dominated by ionic interactions in the oxide and metallic interactions in the metal region. This formalism has been successfully used in the past to investigate oxidation of several metal and alloy surfaces and nanoparticles [25, 26, 27, 28]. To determine the force-field parameters for complex functional forms like CTIP, we require a systematic approach that can enable efficient sampling of most of the available parameter space. We have recently demonstrated that evolutionary global optimization methods such as genetic algorithms (GA) represent a powerful strategy for fitting force fields [29, 30, 31, 32]. Here, we use a combination of GA to perform sampling of the parameter landscape, and local minimization methods (simplex) starting from promising parameter sets identified by GA to obtain an optimized CTIP potential model.

Our paper is organized as follows: Section 2 describes variable charge potential model formalism *i.e.* the CTIP functional forms as well as the training data set, the fitting procedure and parameterization strategies employed by us for potential model development. We also describe the model setup used to investigate early stages of Ta oxidation. Section 3 reports our results on the evolutionary optimization, the optimized CTIP parameters obtained in this study, the various structural and energetic properties predicted by the optimized set of parameters, and their success in describing inter-atomic interactions in $Ta_2O_5$. In Section 4, we apply the newly developed CTIP parameter set to investigate structure and stoichiometry across Ta, Ta-oxide and their hetero-interfaces. Here, we provide a representative example on the application of our newly developed CTIP to study the early stages of oxidation and oxide growth on the (110) surface of body-centered cubic (bcc) Ta at various temperatures. Finally, Section 5 summarizes the key findings and provides concluding remarks.

## 2. Methods

### i. Charge transfer ionic potential (CTIP)

We use the modified charge transfer potential model developed by Zhou *et al.* [23, 24]. This potential model comprises of two parts: ionic interactions in the oxide region are modeled using an electrostatic term ($E_{es}$), and the embedded atom method is used to model metallic interactions i.e. non-electrostatic contributions.

$$E_t = E_{es} + E_m \tag{2.1}$$

In the CTIP model, charge bounds were imposed on cations and anions to prevent them from exceeding their valence charges. This overcomes the limitations of the original Streitz-Mintmire potential [33]. In the modified CTIP, the electrostatic energy is given as below [24]:

$$E_{es} = E_0 + \sum_{i=1}^{N} q_i X_i + \frac{1}{2}\sum_{i=1}^{N}\sum_{j=1}^{N} q_i q_j V_{ij} + \sum_{i=1}^{N} \omega\left(1 - \frac{q_i - q_{min,i}}{|q_i - q_{min,i}|}\right)(q_i - q_{min,i})^2 + \sum_{i=1}^{N} \omega\left(1 - \frac{q_{max,i} - q_i}{|q_i - q_{max,i}|}\right)(q_i - q_{max,i})^2 \qquad (2.2)$$

where $q_{min,i}$ and $q_{max,i}$ are the charge bounds of atom $i$, $q_{min,i} < q_i < q_{max,i}$. Coefficient $\omega$ imposes energy penalty on the metal atoms to gain electrons or lose inner shell electrons and on the oxygen atoms to lose electrons or receive more than two electrons. $X_i$ represents self energy and $V_{ij}$ represents Coulomb interaction [24].

$$X_i = \chi_i + \sum_{j=i_1}^{i_N} k_c Z_j \left([j|f_i] - [f_i|f_j]\right) \qquad (2.3)$$

$$V_{ij} = J_i \delta_{ij} + \sum_{k=j(i_1)}^{j(i_N)} k_c \left([f_i|f_k]\right) \qquad (2.4)$$

In the above equations, $\chi_i$ refer to the electro-negativity and $J_i$ refers to atomic hardness (or self Coulomb repulsion), respectively. As shown in Eqs. (2.5)-(2.7), Coulomb integrals such as $[a|f_b]$ and $[f_a|f_b]$ are calculated assuming atomic charge density distribution for spherical Slater-type orbitals [23]:

$$[a|f_b] = \frac{1}{r_{ab}} - \xi_b \exp(-2\xi_b r_{ab}) - \frac{1}{r_{ab}}\exp(-2\xi_b r_{ab}) \qquad (2.5)$$

$$\text{For } \xi = \xi_a = \xi_b, \quad [f_a|f_b] = \frac{1}{r_{ab}}\left[1 - \left(1 + \frac{11}{8}\xi r_{ab} + \frac{3}{4}\xi^2 r_{ab}^2 + \frac{1}{6}\xi^3 r_{ab}^3\right)\exp(-2\xi r_{ab})\right] \qquad (2.6)$$

$$\text{For } \xi_a \neq \xi_b, \quad [f_a|f_b] = \frac{1}{r_{ab}} - \frac{\xi_a \xi_b^4 \exp(-2\xi_a r_{ab})}{(\xi_a + \xi_b)^2 (\xi_a - \xi_b)^2} - \frac{\xi_b \xi_a^4 \exp(-2\xi_b r_{ab})}{(\xi_a + \xi_b)^2 (\xi_a - \xi_b)^2}$$

$$- \frac{(3\xi_a^2 \xi_b^4 - \xi_b^6)\exp(-2\xi_a r_{ab})}{r_{ab}(\xi_a + \xi_b)^3 (\xi_a - \xi_b)^3} - \frac{(3\xi_b^2 \xi_a^4 - \xi_a^6)\exp(-2\xi_b r_{ab})}{r_{ab}(\xi_a + \xi_b)^3 (\xi_b - \xi_a)^3} \qquad (2.7)$$

Here a=i,j, b=i,j and a≠b. Table 1 lists the optimized charge parameters $\chi_i$, $J_i$, $\xi$ and $Z_i$ as well as the charge bounds for the various elements.

The embedded atom method (EAM) is used to represent the non-electrostatic interactions in the metallic region as follows [23, 33]:

$$E_m = \frac{1}{2}\sum_{i=1}^{N}\sum_{j=i_1}^{i_M}\varphi_{ij}(r_{ij}) + \sum_{i=1}^{N} F_i(\rho_i) \quad (2.8)$$

In Eq. (2.8), $\Phi_{ij}(r_{ij})$ represents the pair-wise interaction energy between atoms $i$ and $j$ that are at distance $r_{ij}$ apart. $N$ is the total number of atoms and $i_M$ is the number of neighbors for atom $i$. For an alloy system, the generalized elemental pair potentials is given by [23]:

$$\phi(r) = \frac{A\exp\left[-\alpha\left(\frac{r}{r_e}-1\right)\right]}{1+\left(\frac{r}{r_e}-\kappa\right)^{20}} - \frac{B\exp\left[-\beta\left(\frac{r}{r_e}-1\right)\right]}{1+\left(\frac{r}{r_e}-\lambda\right)^{20}} \quad (2.9)$$

$F_i$ represents the embedding energy i.e. energy required to embed an atom $i$ into a local site with electron density $\rho_i$ [23]:

$$\rho_i = \sum_{i=1}^{N} f_i(r_{ij}) \quad (2.10)$$

In the above expression, $f_i(r_{ij})$ is the electron density at the site of atom $i$ arising from atom $j$ separated by distance $r_{ij}$, which is given by [23]:

$$f(r) = \frac{f_e \exp\left(-\beta\left(\frac{r}{r_e}-1\right)\right)}{1+\left(\frac{r}{r_e}-\lambda\right)^{20}} \quad (2.11)$$

The embedding energy functions $F$ are selected to work well over a wide electron density range. For a smooth variation of the embedding energy, we fit spline functions across different density ranges [23].

The metal-oxygen and oxygen-oxygen interactions are pairwise and of the same functional form as Eqn. 2.9. In addition to the pair terms, the oxygen interactions also have an embedding energy based on the local electron density. Equation (2.12) gives the functional form of the electron density:

$$f(r) = \frac{f_e \exp\left(-\gamma\left(\frac{r}{r_e}-1\right)\right)}{1+\left(\frac{r}{r_e}-\upsilon\right)^{20}} \quad (2.12)$$

The embedding function is defined as follows [23, 24]:

$$F_j(\rho) = \sum_{i=0}^{3} F_{i,j} \left(\frac{\rho}{\rho_{e,j}} - 1\right)^i \quad ; \quad \rho_{min,j} \leq \rho \leq \rho_{max,j} \tag{2.13}$$

Here, $j$ varies from 0 to M (the number of metals in the alloy oxide; M=1, here). The optimized EAM parameters are listed in Tables 1-5.

**ii. Training data set**

Appropriate training and test datasets, that sufficiently sample the potential energy landscape, are necessary to develop accurate and transferable force fields. To train our CTIP model, we employ three experimentally reported crystalline polymorphs of $Ta_2O_5$ (see Fig. 1; Ref. [34, 35]), namely, monoclinic (*C2/c*), *δ*-hexagonal (*P6/mmm*) and *β*-orthorhombic (*Pmmm*). For each of these polymorphs, the lattice constants, internal coordinates, cohesive energy, and equation of state (energy vs. volume) from DFT+*U* calculations are included in the training set. In addition, the training set also consists of 13 independent elastic constants of the monoclinic phase. Surface energies of bcc Ta and monoclinic $Ta_2O_5$, the bulk modulus of all three polymorphs, and the structure and dynamics of the polymorphs at 300 K are not used in fitting but used for cross validation tests of the interatomic potential parameters.

While experimental data can also be used in the evolutionary optimization and cross validation tests, there is insufficient experimental data on cohesive energies and elastic constants of different polymorphs of $Ta_2O_5$ to develop an extensive training set. The available DFT calculations in literature also have a range of reported values based on the exchange correlation functional and additional approximations. Hence, for consistency, we derived the values of these datasets directly from DFT calculations performed with the generalized gradient approximation (GGA) as parameterized by Perdew-Burke-Ernzerhof (PBE) [35], along with a Hubbard *U* correction [34, 36, 37] on the Ta 5*d* electrons. The projector-augmented wave formalism as implemented in the Vienna *Ab-initio* Simulation Package (VASP) is used. We employ a plane wave energy cutoff of 500 eV and sample the Brillouin Zone (BZ) using Γ-centered Monkhorst-Pack k-point grids, of 2×6×4 for the monoclinic, 3×3×6 for the *δ*-hexagonal, and 6×6×4 for the *β*-orthorhombic structures. These particular grids were selected based on tests done for different BZ samplings using from 1 to 123 irreducible k-points. The chosen k-point grids (see supporting information Fig. SM5) yield cohesive energies that are within 1 meV/u.f. compared to those obtained with larger grid sizes. For the Hubbard *U*-correction, we choose $U = 1.35$ eV as derived from linear response by Ivanov *et al.* [34, 36]. For the structural optimization, and the subsequent calculation of cohesive energy, equation of state, elastic constants, and surface energies, the atoms were relaxed until the forces on each atom was less than $10^{-4}$ eV/Å.

### iii. Evolutionary strategy to optimize the CTIP parameters

Single metal oxides under the CTIP formalism require 50 parameters to be appropriately trained. The 24 EAM parameters for the metal-metal interactions in Ta have been previously optimized by Zhou et al. [38]. We adopt these parameters to describe the Ta-Ta interactions, and list them in Table 1. The remaining 26 parameters are appropriately trained using an evolutionary strategy.

Interatomic potential parameterization involves minimizing the objective function ($\Delta$), which is a measure of the error between the training set and the fit. For a 26-dimensional parameter set, the objective function landscape can be complex with multiple minima. Using local optimization strategies alone, the optimized parameters obtained will depend heavily on the initial guess as the objective settles into the nearest local minimum (e.g., see Fig. 1e for a schematic representation of the objective landscape of a 2-dimensional parameter set). An evolutionary strategy such as genetic algorithm (GA) based global optimization offer a more efficient sampling of the parameter space. Such an approach has been successfully used for fitting force fields for a wide class of materials systems [29, 30, 31, 32].

Here, we start with a random population of $N_p = 50$ parameter sets with 26 parameters each. Appropriate genetic operations such as crossover and mutation are performed to obtain derived (child) parameter sets [39]. The fitness of a particular parameter set (individual) is assessed by the accuracy of its predicted properties against the training set, i.e. $\Delta$ is defined as a weighted sum of errors in predicted properties (computed using MD package LAMMPS [40]). The value of $\Delta$ is evaluated for each individual and $N_p$ individuals with the lowest objective function values are chosen for the next iteration of genetic operations.

Once the GA run converges, additional local optimization is performed starting from promising parameter sets (15 sets with the lowest error in prediction) using the Simplex algorithm. Test set validations are performed before deciding on the final set of optimized parameters (see Fig. 1 for the schematic representation of the evolutionary strategy used in this work). The optimized CTIP parameters are listed in Tables 2-5.

## iv. Oxidation Simulation Set-up

The optimized parameters are used to investigate the initial stages of oxidation of metallic Ta by employing a similar approach used by Sankaranarayanan et al. [25, 26, 27]. The LAMMPS MD package [40] is used to perform all the simulations in this work. Two pristine Ta (110) surfaces $4.67 \times 4.67\ nm^2$ are generated for the oxidation simulation by introducing vacuum slabs of 5.63 nm on each side of the metal substrate, along a direction normal to the surfaces (see Fig. 2). Periodic boundary conditions are applied in the plane of the surface. A sufficiently long cutoff (12 Å) is chosen for the short-range Coulomb interactions. Long-range Ewald summations [24] for the Coulomb interaction are performed only along the periodic directions.

The slabs are thermalized by annealing in the temperature range 0 K to 300 K in steps of 50 K using a Nose-Hoover thermostat [41]. For each temperature, MD simulations are performed in the NVT ensemble for 5 ps. This first round of thermalization is performed by ignoring the dynamic charge transfer between tantalum atoms. This is appropriate since the charges are zero for a pure metallic system. An additional equilibration run of 100 ps is performed at the desired temperature by including the charge dynamics and allowing the box to relax in the $x$- and $y$- directions by using a Nose-Hoover barostat [42]. As expected, we find that the atomic charges fluctuate around a zero value in the pure metal with a magnitude of $\pm\ 0.05e$ at the two outer layers and of $\pm\ 0.01e$ in the bulk.

The oxidation of the metal substrates is initiated by introducing either $O_2$ molecules or atomic oxygen (in separate simulations) in the vacuum slab at at random *x*, *y* and *z* positions. The velocities of the $O_2$ or O are chosen from a Maxwell-Boltzmann distribution corresponding to the desired temperature ranging from 300-900 K. Additionally, reflecting boundary conditions are imposed on the molecules that might reach the *z*-direction simulation box boundaries. The gas pressure is maintained constant during the simulation by introducing a new $O_2$ molecule or atomic oxygen only when the previous molecule dissociates and forms bonds with the metal atoms. This is achieved by tracking the planar position of the growing oxide-gas interface (calculated based on the average positions of the outermost metal ions). A new oxygen atom/molecule is inserted only when a previous one bonds with the metal ion at the oxide-gas interface and enters into the oxide-gas interface ($z_{oxygen}$ <= $z_{oxide-gas\ interface}$). The equations of motion are integrated using a velocity Verlet scheme with time steps of 0.5 fs for a total time of 200 ps for each simulation. The charge relaxation procedure used to minimize the electrostatic energy subject to the electro-neutrality principle is performed every MD step.

## 3. Results and Discussion

### i. Optimization of CTIP parameters for Ta/Ta oxide

Using a single-objective evolutionary optimization technique detailed in Fig. 1, we determine the EAM+QEq (CTIP) parameters for a Ta-O binary system by fitting against structural, thermodynamic, and elastic properties of the three main polymorphs of $Ta_2O_5$, namely, monoclinic (*C2/c*), hexagonal (*P6/mmm*) and orthorhombic (*Pmmm*). During each round of optimization, several 26-parameter space 50-population GA runs are iterated for 100 generations (see pseudo-code in supporting information for GA optimization [39]). This allows for a more detailed sampling of the parameter space. Fifteen of the best sets from each round of optimization are used to further fine-tune the parameter ranges for the next round of optimization. This is continued till we obtained the optimized parameters, reported here, that have the lowest objectives with the training and the test sets. The optimized EAM+QEq parameters are listed in Tables 2-5.

**Table 1. EAM parameters for metal-metal interaction**

| Metal | $r_e$(Å) | $f_e$ | $\rho_e$ | $\rho_s$ | α | β | A (eV) |
|---|---|---|---|---|---|---|---|
| Ta | 2.860082 | 3.086341 | 33.787168 | 33.787168 | 8.489528 | 4.527748 | 0.611679 |

| Metal | B(eV) | κ | λ | $F_{n0}$ (eV) | $F_{n1}$ (eV) | $F_{n2}$(eV) | $F_{n3}$ (eV) |
|---|---|---|---|---|---|---|---|
| Ta | 1.032101 | 0.176977 | 0.353954 | -5.103845 | -0.405524 | 1.112997 | -3.585325 |

| Metal | $F_0$(eV) | $F_1$(eV) | $F_2$(eV) | $F_{3+}$(eV) | $F_{3-}$(eV) | η(eV) | $F_e$(eV) |
|---|---|---|---|---|---|---|---|
| Ta | -5.14 | 0.00 | 1.640098 | 0.221375 | 0.221375 | 0.848843 | -5.141526 |

**Table 2. CTIP parameters for the simulated elements**

| Element | $q_{min}$ | $q_{max}$ | χ (eV) | J (eV) | ξ(Å$^{-1}$) | Z(e) |
|---|---|---|---|---|---|---|
| O | -2.00 | 0.00 | 5.481730 | 15.128200 | 2.143957 | 0.000000 |
| Ta | 0.00 | 5.00 | 0.000000 | 10.758300 | 1.036140 | 1.159850 |

**Table 3. Optimized EAM parameters for pair potentials**

| Pair | $r_e$(Å) | A | β | A (eV) | B (eV) | κ | λ |
|---|---|---|---|---|---|---|---|
| O-Ta | 2.088696 | 6.295628 | 1.864984 | 1.301541 | 2.475937 | 0.488616 | 0.262364 |
| O-O | 3.668766 | 5.931086 | 4.104248 | 0.415946 | 0.706278 | 0.240129 | 0.753456 |

**Table 4. EAM parameters for oxygen electron density function**

| $f_e$ | γ | ν |
|---|---|---|
| 1.888839 | 2.708917 | 0.694904 |

**Table 5. Optimized EAM parameters for oxygen embedding energy spline function**

| i | $F_{0,i}$(eV) | $F_{1,i}$(eV) | $F_{2,i}$(eV) | $F_{3,i}$(eV) | $\rho_{e,i}$ | $\rho_{min,i}$ | $\rho_{max,i}$ |
|---|---|---|---|---|---|---|---|
| 0 | -1.576011 | -1.757687 | 1.212659 | 1.394335 | 63.631547 | 0 | 63.631547 |
| 1 | -1.866573 | -1.806293 | 0.871914 | 0.000000 | 74.860644 | 63.631547 | ∞ |

## ii. Performance of the Newly Developed CTIP parameters

Next, we assess the accuracy of our newly developed EAM+QEq parameters by comparing our predicted properties against the training and test sets. The fitted and test set properties of the three polymorphs are compared against experiments and first-principles calculations in Tables 6-10. The monoclinic structure, energy and elastic constant components (also see supporting information Fig. SM2) are captured with reasonable accuracy, and so are the structures and energies of the hexagonal and orthorhombic polymorphs. The energy ordering between the three polymorphs is also accurately computed by the CTIP parameters. In Fig. 3 we show the equation of state (energy vs. volume) for all the polymorphs along with the corresponding Murnaghan fit [43] used to compute the bulk modulus, using both CTIP and DFT calculations. We also report the surface energies of monoclinic $Ta_2O_5$ for the (001), (010) and (100) surfaces (see Fig. 4 for surface configurations). The predicted surface energies are compared against DFT values in Table 9. Finally, we also report bcc Ta surface energies and find excellent agreement between the predicted values, first-principles calculations and experiments (see Table 10).

Table 6. Properties of monoclinic $Ta_2O_5$ given by the EAM+QEq (CTIP) interatomic potential parameters developed in this work in comparison with experiments and first-principles calculations.

| Monoclinic | | | | |
|---|---|---|---|---|
| Properties | Expt. [44] | CTIP | DFT-GGA+U | DFT-GGA [45] |
| a (Å) | 12.79 | 12.93 | 12.93 | 15.07 |
| b (Å) | 4.85 | 4.71 | 4.93 | 4.93 |
| c (Å) | 5.53 | 5.53 | 5.59 | 5.60 |
| α | 90 | 90 | 90 | 90 |
| β | 104.26 | 103.58 | 103.24 | 123.19 |
| γ | 90 | 90 | 90 | 90 |
| $E_c$ (eV/u.f.) | - | -64.53 | -64.49 | -68.31 |
| B (GPa) | - | 196 | 122 | 137 |
| $C_{11}$ (GPa) | - | 263 | 282 | 208 |
| $C_{22}$ (GPa) | - | 297 | 215 | 173 |
| $C_{33}$ (GPa) | - | 230 | 239 | 272 |
| $C_{44}$ (GPa) | - | 90 | 126 | 83 |
| $C_{55}$ (GPa) | - | 52 | 56 | 107 |
| $C_{66}$ (GPa) | - | 32 | 57 | 68 |
| $C_{12}$ (GPa) | - | 132 | 81 | 63 |
| $C_{13}$ (GPa) | - | 129 | 88 | 112 |
| $C_{15}$ (GPa) | - | -8 | -40 | 0 |
| $C_{23}$ (GPa) | - | 152 | 103 | 114 |
| $C_{25}$ (GPa) | - | -2 | -22 | 0 |
| $C_{35}$ (GPa) | - | -29 | -25 | 0 |
| $C_{46}$ (GPa) | - | 8 | 0 | 0 |

**Table 7. Properties of δ-hexagonal $Ta_2O_5$ given by the EAM+QEq (CTIP) interatomic potential parameters developed in this work in comparison with experiments and first-principles calculations.**

| Hexagonal | | | | |
|---|---|---|---|---|
| Properties | Expt. [46, 47, 48] | CTIP | DFT-GGA+U | DFT [34, 46] |
| a (Å) | 7.25 - 7.34 | 7.11 | 7.33 | 7.12 - 7.32 |
| b (Å) | 7.25 - 7.34 | 7.11 | 7.33 | 7.12 - 7.32 |
| c (Å) | 3.88 | 3.96 | 3.89 | 3.83 - 3.88 |
| α | 90 | 90 | 90 | 90 |
| β | 90 | 90 | 90 | 90 |
| γ | 120 | 120 | 120 | 120 |
| $E_c$ (eV/u.f.) | - | -63.48 | -62.63 | -59.84, -60.34 |
| B (GPa) | - | 268 | 217 | - |

**Table 8. Properties of β-orthorhombic $Ta_2O_5$ given by the EAM+QEq (CTIP) interatomic potential parameters developed in this work in comparison with experiments and first-principles calculations.**

| Orthorhombic | | | | |
|---|---|---|---|---|
| Properties | Expt. [49, 50] | CTIP | DFT-GGA+U | DFT [35, 45, 51, 52] |
| a (Å) | 3.68 | 3.62 | 3.69 | 3.55 - 3.75 |
| b (Å) | 3.90 | 3.93 | 3.89 | 3.75 - 3.89 |
| c (Å) | 6.23 | 6.24 | 6.52 | 6.53 – 7.9 |
| $E_c$ (eV/u.f.) | - | -62.26 | -62.30 | -66.14 |
| $\Delta E – \beta-\delta$ (eV/u.f.) | - | 1.22 | 0.33 | 0.4 |
| B (GPa) | - | 267 | 197 | - |

Table 9. Surface energies in J/m$^2$ of monoclinic Ta$_2$O$_5$.

| Surface | CTIP | DFT |
|---|---|---|
| (100) | 5.32 | 4.92 |
| (010) | 2.19 | 1.06 |
| (001) | 3.72 | 2.09 |

Table 10. Surface energies in J/m$^2$ of bcc Ta compared against experiments, first-principles and reported MD calculations. Though Zhou et al.'s EAM parameters for Ta (used in CTIP) were not explicitly trained against the surface energies, they are captured with reasonable accuracy.

| Surface | CTIP | EAM [53] | DFT [45] | Expt. [54] |
|---|---|---|---|---|
| (110) | 1.96 | 2.29 | 2.34 | 2.49 |
| (100) | 2.29 | 2.75 | 2.47 | - |
| (111) | 2.31 | 2.98 | 2.70 | - |

## 4. Case study: Oxidation of Ta (110) surface

We apply this newly developed force field to study the oxidation of a Ta (110) surface in the presence of atomic and molecular oxygen. MD simulations are performed up to 200 ps for temperatures ranging from 300-900 K. Structure and stoichiometry information in the metal/oxide/gas interface is extracted from the MD simulation and used to gain insights into the evolution and morphology of the growing oxide film.

Fig. 5 shows the oxidation kinetics curves for various temperatures. As expected, we see that the oxygen adsorption increases with temperature and that atomic oxygen is more active than molecular oxygen (Fig. 5 a, d). In addition, we observe that the oxide film thickness (see Fig. 5 b, e) reach a limiting value of ~0.7-0.9 nm. Similar self-limiting thicknesses have been obtained for the natural oxidation of Zr both from MD and experiments [25, 27, 55]. In addition, Mathieu et al. [56] reported natural oxide film thicknesses ranging from 0.2 – 5 nm for several metal oxides (Al, Fe, Ni) via Auger Electron Spectroscopy (AES) and x-ray photoelectron spectroscopy (XPS). However, experimental reports on self-limiting thickness of the oxide film during natural oxidation of Ta are scarce. Most experimental work deals with anodic oxidation and plasma-enhanced chemical vapor-deposition to deposit films over 100 nm thick [57, 58]. Available experimental data for natural oxidation have a wide variability in the reported film thicknesses – 0.7 nm via AES [56], 1.9 nm via XPS [56], and 3.0 nm via XPS [59]. Furthermore, these measurements are for longer oxidation times much beyond 100 s. Such time scales are not accessible to the current MD simulations, making direct comparison challeging. With regard to the atomic oxidation of Ta, to the best of our knowledge, there is no available experimental data in the simulated low-temperature range. However, the enhancement in oxidation kinetics observed in our simulations agrees well with previous reports for low-temperature atomic oxidation of Ag and Si (110) surfaces [60, 61, 62].

Carbera-Mott theory for oxidation kinetics at low temperature in ultrathin films [63, 64, 65] may be utilized to analyze the oxidation kinetics curves (Fig. 5) and used to estimate of the activation energy barrier for oxidation on Ta surfaces using O and $O_2$. According to this theory, the driving force for oxidation is an induced internal electric field that drives the ionic transport, which accelerates the initial oxidation but is rapidly attenuated with increasing oxide film thickness. Such a model predicts a logarithmic growth rate for metal oxides. The expression for the rate equation is given as follows [25]:

$$\frac{dL}{dt} = C \, exp\left(-\frac{W_0 - \frac{1}{2}qaE + \lambda L}{k_B T}\right) \qquad (3.14)$$

where, $L$ is the film thickness, $W_0$ represents the intrinsic barrier for ionic jumps between two positions in the oxide film, $q$ is the charge on the ion, $2a$ is the jump length, $E$ is the electric field, $\lambda$ is the structure term, $T$ is the temperature, $k_B$ is the Boltzmann constant, and $C$ is a constant. The solution to the above equation yields a direct logarithmic growth law given by [25]:

$$L(t) = \left(\frac{k_B T}{\lambda}\right) ln[1 + \mu(T)t] \qquad (3.15)$$

$$\text{where } \mu(T) = \left(\frac{\lambda}{k_B T}\right) C \, exp\left(-\frac{W_0 - \frac{1}{2}qaE}{k_B T}\right) \qquad (3.16)$$

We fit the simulated film thickness data to $L(t) = A \ln(1 + Bt)$ to obtain estimates for the structure term, $\lambda = \frac{k_B T}{A}$ and $\mu(T) = B$. The structure term $\lambda/k_B$ varies linearly with temperature if logarithmic growth law is valid. We observe the expected linear dependence for the temperature range 300-600 K for both natural and atomic oxidation (see Fig. 5 c, f). In addition, we see that atomic oxygen has lower values for structure term indicating lower energy barrier or faster kinetics for atomic oxidation than natural oxidation. Interestingly, we observe a deviation from the linear dependence in for $\lambda$ vs. $T$ at 900 K (see supporting information Fig. SM3). Such a deviation from logarithmic kinetics of Carbera-Mott theory has been observed for the oxidation of Ta above ~320°C [18, 66]. While the exact mechanism of high temperature oxidation is not clear from the current MD simulations, it is anticipated to have parabolic kinetics [18]. Longer simulation timescales than currently accessible from the current MD simulations is necessary to investigate this in detail and is subject of future studies.

Oxide growth in metals can proceed via the following three mechanisms: (a) the metal ions alone migrate and the new oxide forms at the oxide/gas interface; (b) oxygen ions alone migrate and the new oxide forms at the oxide/metal interface; and (c) both the ions migrate and the new oxide layer can form at both interfaces and/or within the existing oxide [57]. It has been experimentally observed that the transport numbers for oxygen ions is thrice that of tantalum ions during oxide growth in the metal and that the oxide thickens through the formation of new oxide at both interfaces [57]. Hence, the kinetics of the process is limited by cation transport. An Arrhenius fit to Eqn. 3.16 can be used to estimate the overall energy barrier $W_0 - \frac{1}{2}qaE$. We find this to be $0.13 \pm 0.02$ eV for molecular/natural oxidation and $0.043 \pm 0.015$ eV for atomic oxidation, respectively. If the value of the induced internal electric field that drives the ionic transport ($E$) is known, we can get an exact estimate of the intrinsic energy barrier $W_0$. This electric field lowers the energy barriers for the outward migration of cations in the developing oxide film. The induced internal electric field, $E$, has been experimentally estimated to be in the range 4-7 MV/cm for Ta [67]. Other estimates for the Mott potential ($V_K$) for the low temperature oxidation of Ta are in the 0.5-0.65 V range at 523 K [64]. For an ~0.8 nm thick oxide film (as obtained in this work at 500 K), we calculate $E = \frac{V_K}{L}$ in the range 6.25-8.13 MV/cm. We use $E = 6.25$ MV/cm in our calculations here. This is similar to values used by Sankaranarayan et al.'s prior MD studies on oxidation of Zr and Al [25, 26, 27]. Using an estimated charge ($q$) of $4.65 \pm 0.18e$ for the charge of Ta in $Ta_2O_5$ and a cation jump distance ($a$) of $3.14 \pm 0.07$Å for molecular oxidation and $3.13 \pm 0.08$Å for atomic oxidation (estimated from the first peak distance in the Ta-Ta radial distribution function within the oxide film), we obtain $W_0 = 0.59 \pm 0.05\ eV$ for molecular oxidation and $W_0 = 0.49 \pm 0.04\ eV$ for atomic oxidation. Experimental estimates for energy barriers for molecular oxidation of Ta for temperatures lower than ~320°C are 0.54 eV [66], 0.62 eV for polycrystalline samples [68] and 1.2 eV for single crystal Ta (100) surface [68].

In Fig. 6, we show the atomistic representation of the time evolution of the oxidation process at a representative temperature of 600 K. We see that the oxide layer is amorphous and that the oxygen density on the substrate surface becomes more or less uniform within 200 ps. To further elucidate the structural characteristics of the oxide scale, we analyze the oxide structure using partial pair distribution functions (PDF) and bond-angle distributions (ADF) by averaging over 20 trajectories between 195 ps and 200 ps of the oxidation simulation. Fig. 7(a) and 7(c) show the Ta−O PDF in the oxide scale interior at various temperatures for natural and atomic oxidation, respectively. The position of the first peak in $g_{TaO}(r)$ gives the Ta−O bond length to be around 2.15 Å at all temperatures and for both molecular and atomic oxidation. This is consistent with Ta-O distances observed in amorphous $Ta_2O_5$ films under typical deposition conditions [51]. Fig. 7(b) and 7(d) show O−Ta−O bond-angle distribution in the interior of the oxide scale at various temperatures for Ta oxidation using $O_2$ and O, respectively. For all cases, we find that the bond-angle distribution is spread over a broad range between 50° and 180° (indicative of the amorphous nature of the oxide scale). Additionally, we observe two distinct peaks at 70° and 135° in the bond-angle distribution throughout the oxide structure. This is typical of an amorphous and non-stoichiometric oxide scale [27]. The ADF is relatively unaffected by temperature and atomic/molecular nature of the oxidizing species.

We further analyze the time evolution of the oxide film composition (average O/Ta stoichiometry ratio) for both natural and atomic oxidation at different temperatures spanning 300−900 K (Fig. 8 and supporting information Fig. SM4). The 2-d color-maps in Fig. 8 have time (in ps) along the horizontal-axis, the spatial coordinate normal to the metal substrate surface along the vertical-axis, and is colored by the O/Ta stoichiometry ratio. We see that the oxide film thickness increases with temperature and that the O/Ta ratio (and consequently the charge distribution) shows a gradient with maximum stoichiometry ratio at the oxide/gas interface and minimum O/Ta ratio at the oxide/metal interface. Additionally, the O/Ta ratio of the self-limiting oxide film increases with temperature (see supporting information Fig. SM4). For atomic oxidation, the O/Ta ratio approaches 2.1-2.3 within 200 ps for temperatures from 300-900 K. On the other hand, the corresponding O/Ta ratio for molecular oxidation is significantly lower than the stoichiometric ratio and varies from approximately 1.48 at 300 K to 2.11 at 900 K (within 200 ps of oxidation). This lower O/Ta ratio is attributed to slower oxidation kinetics leading to O-deficient oxide films. Additionally, the O/Ta ratio is also lower than 2.5, which is the stoichiometric ratio in crystalline tantalum oxide. This further hints at the amorphous nature of the oxide scale.

## 5. Conclusions

We have developed the first variable-charge potential for $TaO_x$ material systems and used it to study Ta oxidation by both molecular ($O_2$) and atomic (O) oxygen *via* molecular dynamics simulations. We adopt the charge-transfer ionic potential (CTIP) formalism to treat charge transfer among atoms and investigate the oxidation kinetics during the initial stages of tantalum oxide growth. Oxidation is studied as a function of both temperature and the atomic/molecular nature of the oxidizing species. We report intrinsic activation barriers of 0.59 eV and 0.49 eV for natural and atomic oxidation, respectively. The lower activation energy barrier for atomic oxidation is likely responsible for the observed increase in the oxide growth kinetics. We further characterize the structure and morphology of the oxide films formed during natural and atomic oxidation of Ta. Structural analysis reveals self-limiting thicknesses in the range of ~0.7-0.9 nm for the amorphous oxide film obtained after 200 ps of simulation time at various temperatures. We also report a gradient in the O/Ta ratio (and consequently the charge distribution) through the oxide film, with maximum at the oxide/gas interface and minimum O/Ta ratio at the oxide/metal interface. Our findings are in good agreement with previous experiments on low-temperature Ta oxidation [56, 66, 68]. Furthermore, this CTIP potential is suitable for investigating the atomistic mechanisms responsible for the sub-nanosecond timescale switching behavior of TaOx memristors, and other transport phenomena at Ta/TaO$_x$ interfaces under external stimuli.

**Acknowledgement**

This research used resources of the National Energy Research Scientific Computing Center, a DOE Office of Science User Facility supported by the Office of Science of the U.S. Department of Energy under Contract No. DE-AC02-05CH11231. Use of the Center for Nanoscale Materials was supported by the U. S. Department of Energy, Office of Science, Office of Basic Energy Sciences, under Contract No. DE-AC02-06CH11357.

**Supporting Information**

Additional data comparing 1,2) equations of state from DFT *vs*. MD, 3) oxidation kinetic curve, 4) oxygen stoichiometry variation as a function of temperature, 5) DFT k-point convergence for all polymorphs, and 6) pseudo-code for the genetic algorithm employed in this work.

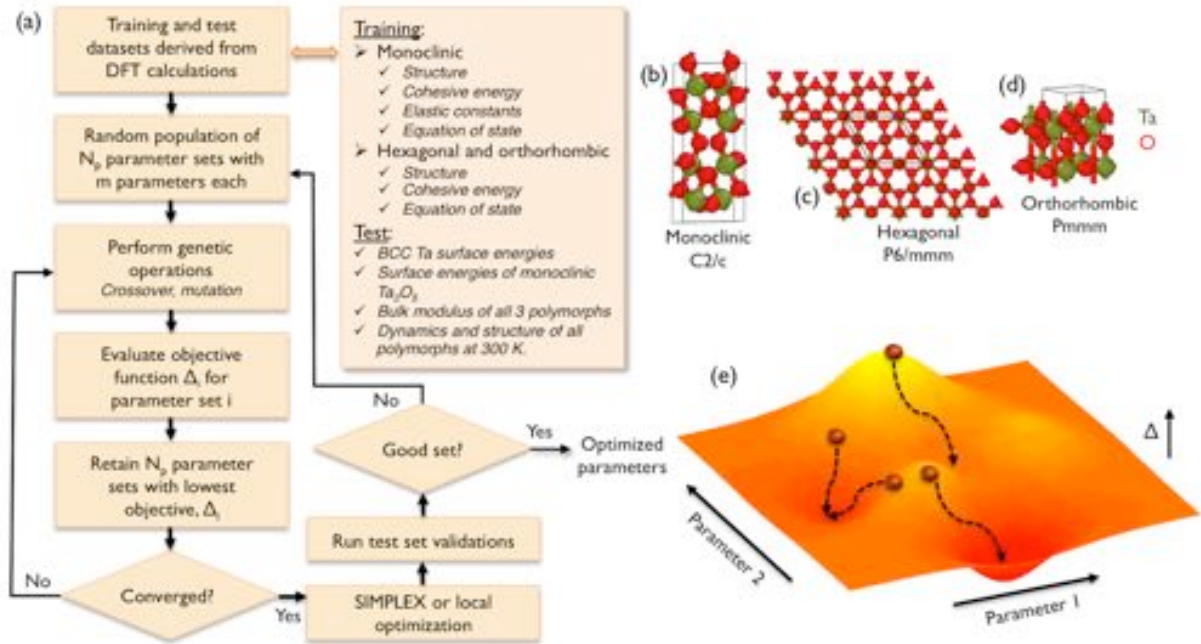

**Figure 1:** (a) Schematic representation of the two-stage evolutionary optimization strategy employed here to parameterize the charge transfer interatomic potential for tantalum oxide. The structures, cohesive energies and energetic ordering of three polymorphs of $Ta_2O_5$, namely, (b) monoclinic, (c) hexagonal and (d) orthorhombic are included in the fit. (e) Schematic representation of the optimization for a sample 2-dimensional problem. The optimized parameters obtained will depend heavily on the initial guess as the objective settles into an appropriate local minimum. An evolutionary algorithm/approach provide a route for a more efficient sampling of a multidimensional parameter space.

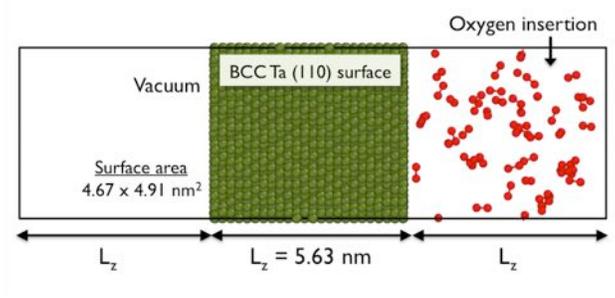

**Figure 2:** Schematic showing the simulation set-up for the oxidation case study: unit cell of substrate and the vacuum slabs surrounding it. Note that the vacuum-facing Ta surfaces are (110).

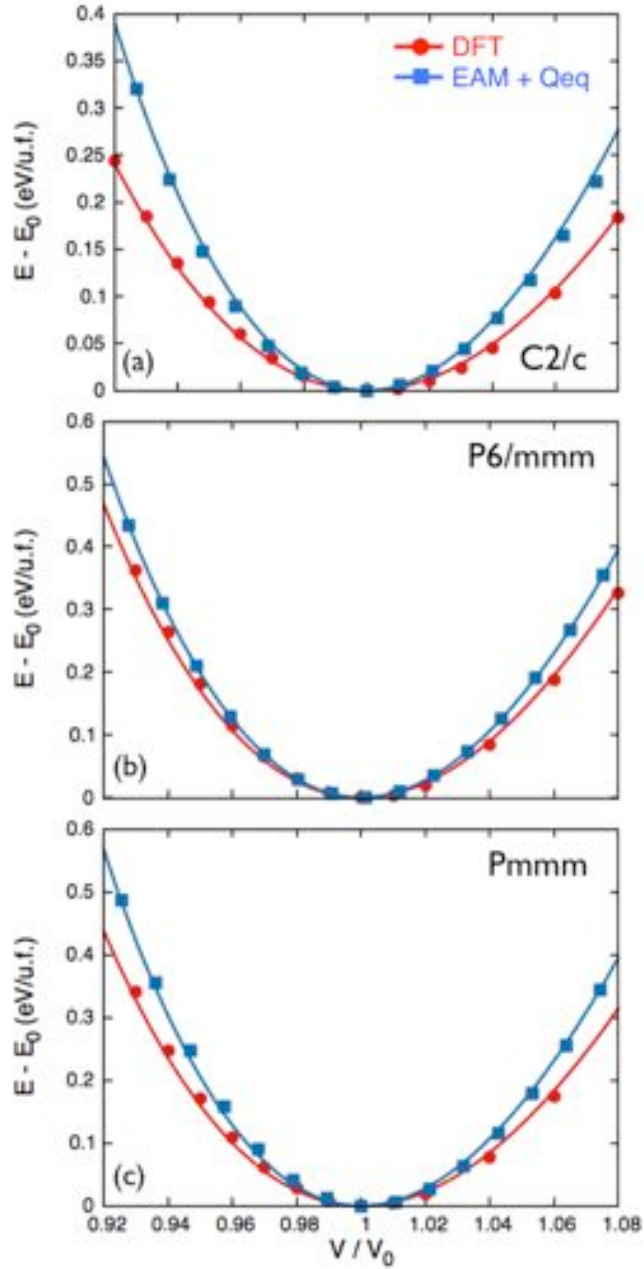

**Figure 3:** Comparison of the equation of state (EOS) near equilibrium predicted by our CTIP-EAM model with our DFT calculations. The equations of state for (a) monoclinic, (b) hexagonal and (c) orthorhombic $Ta_2O_5$ calculated using the EAM+Qeq parameters developed in this study (solid blue squares) and DFT calculations (solid red circles). The solid lines correspond to the Murnaghan fit. The energies are relative to the cohesive energy of the crystal at equilibrium ($E_0$) as evaluated by the corresponding level of theory. The crystal volumes are normalized by the equilibrium value in the framework of the corresponding level of theory.

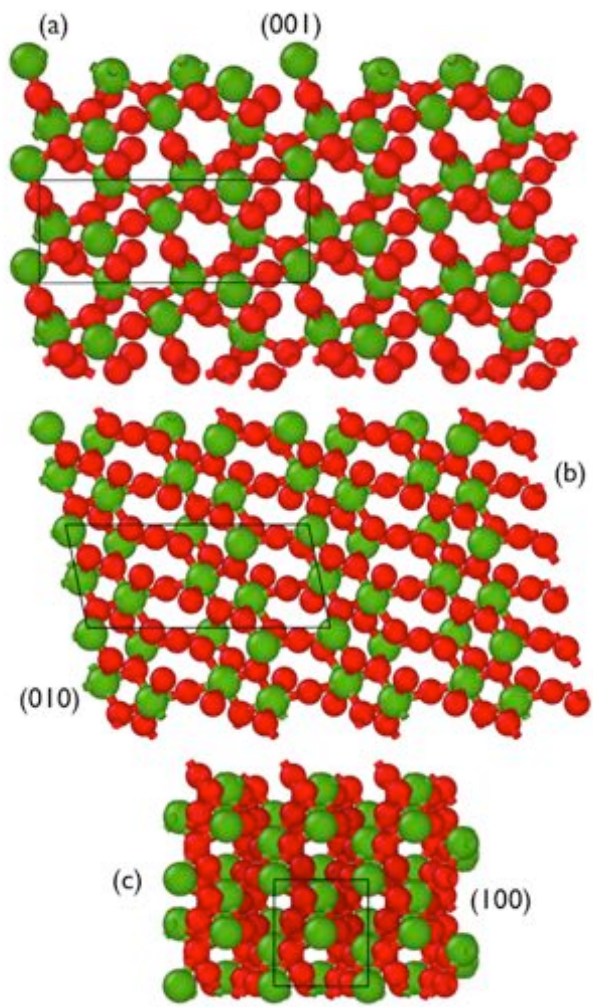

**Figure 4:** Surface configurations of monoclinic Ta$_2$O$_5$ used in our test calculations. For clarity, the surface unit cell (whose edges are shown in black) is repeated as appropriate along the principal directions in the plane of the surface. The (001) and (100) surfaces are O-terminated, while the (010) surface is Ta-terminated. Ta atoms are shown by green spheres and oxygen by red.

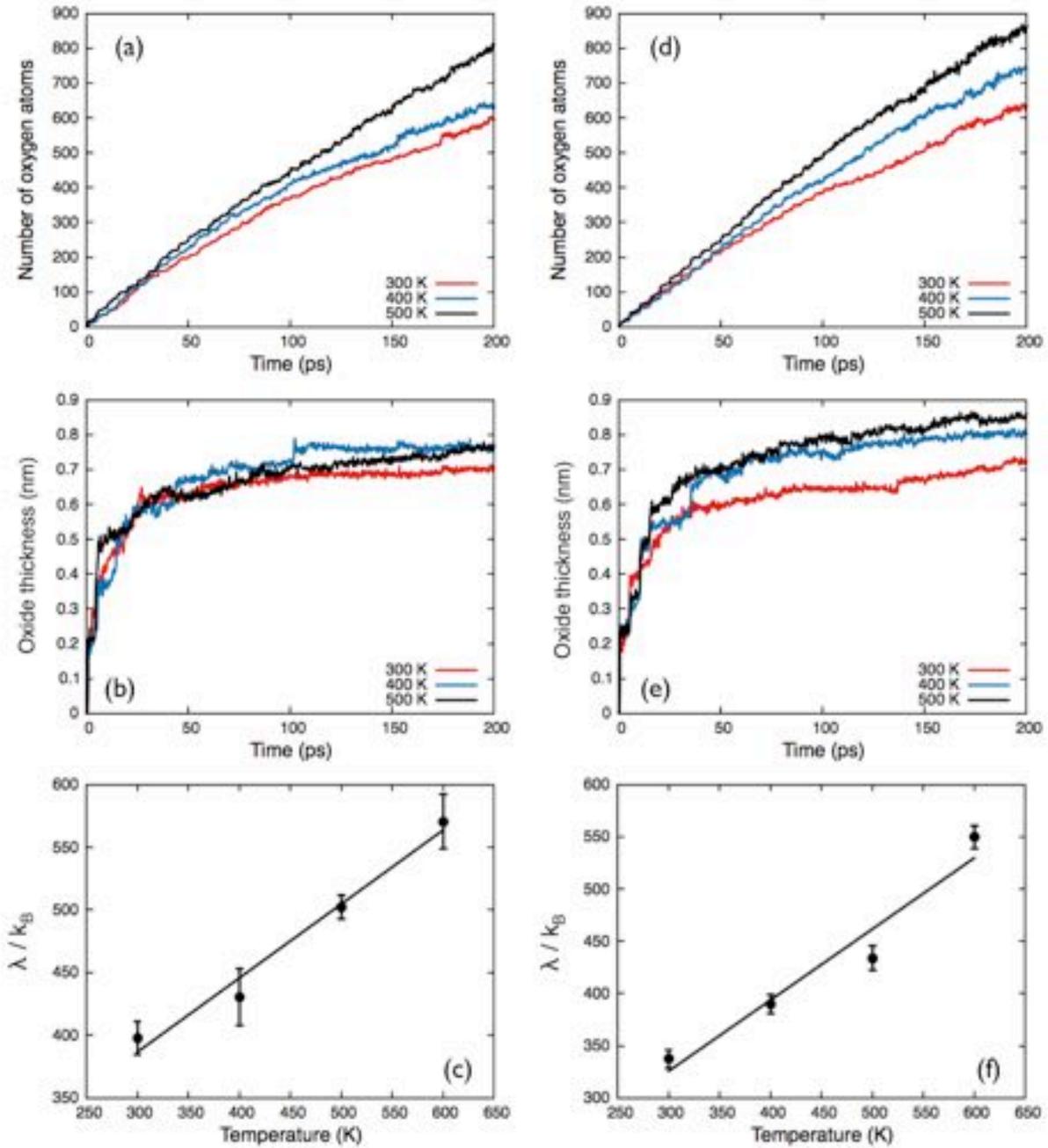

**Figure 5:** Oxidation kinetics curves of Ta(110) as a function of temperature. (a), (b) and (c) show the number of adsorbed O atoms, oxide film thickness and the structure term obtained by assuming logarithmic kinetics, respectively, for molecular/natural oxidation. (d), (e) and (f) are the corresponding plots for atomic oxidation.

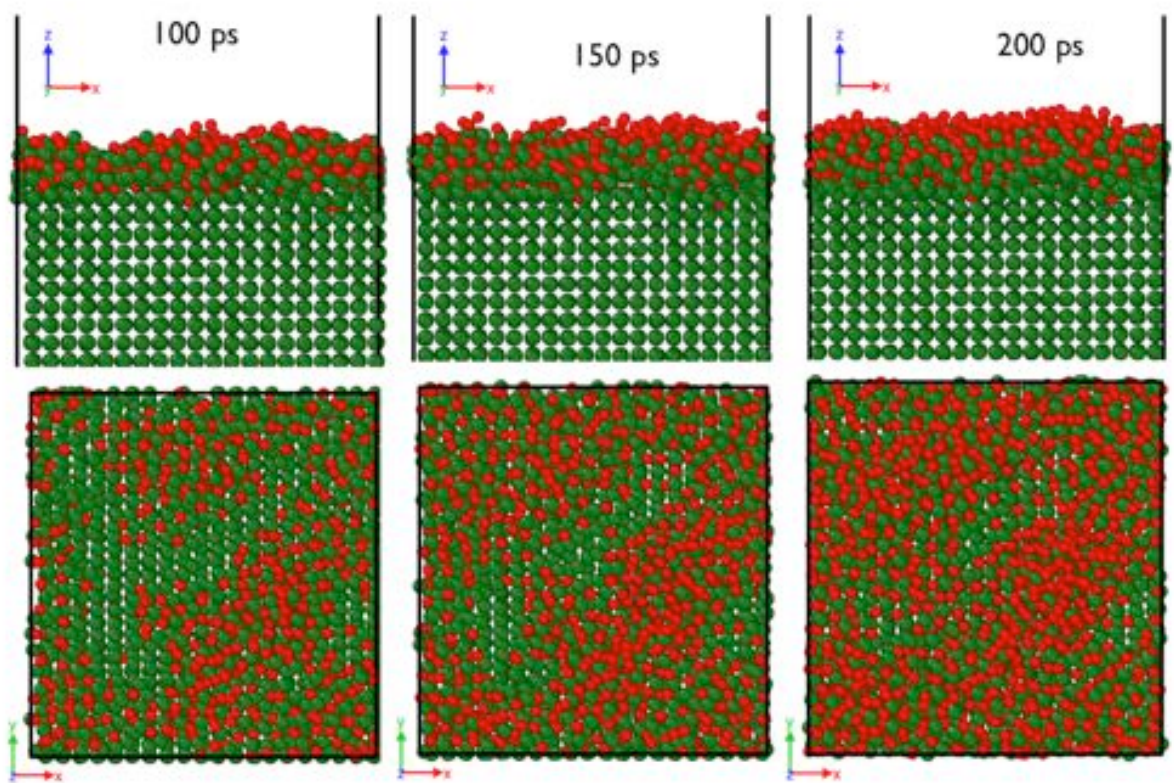

**Figure 6:** Atomistic representation of the time evolution of the molecular oxidation process at representative temperature of 600 K. **Top panel**: Side view; **Bottom panel**: Top view of the Ta (110) surface. Ta atoms are shown by green spheres and oxygen by red. The oxygen molecules in the gaseous phase have been removed from the depiction above.

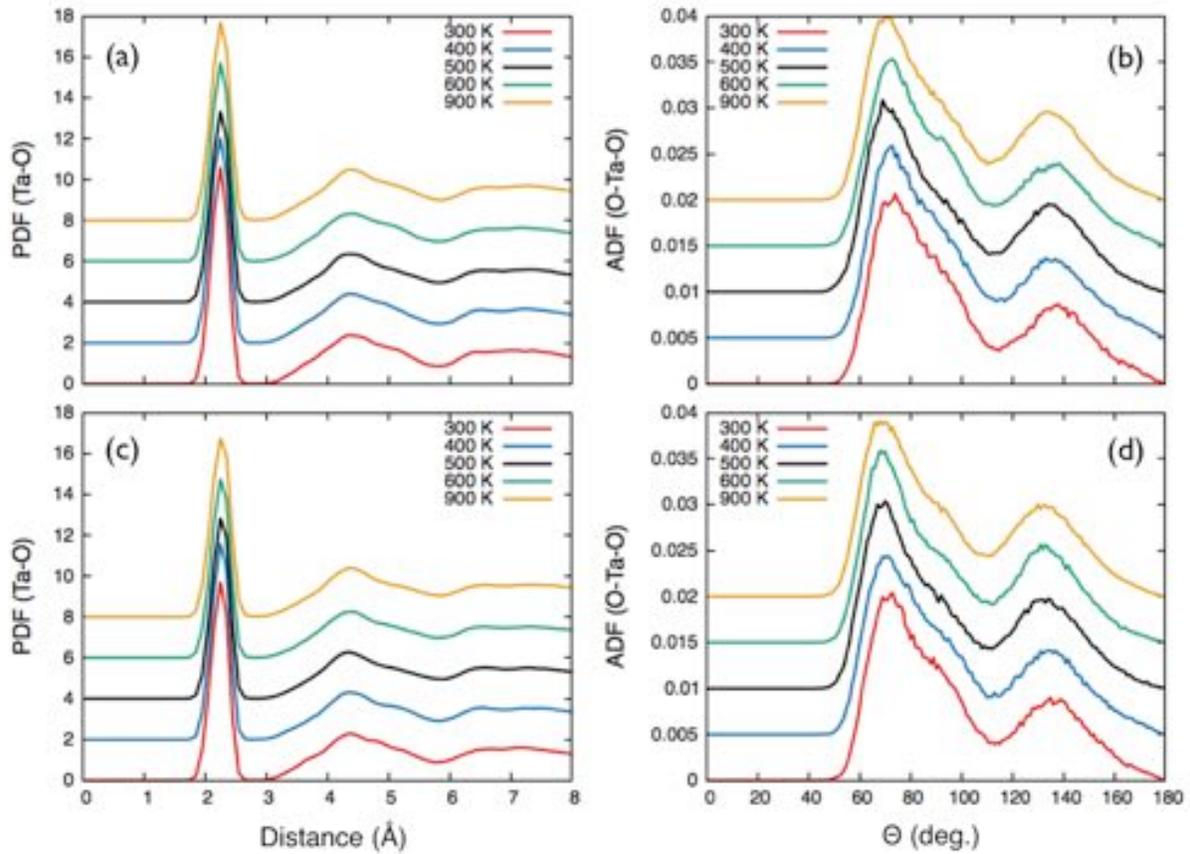

**Figure 7:** Structural evolution of the oxide films as a function of temperature. (a) Ta-O radial distribution function (RDF), and (b) O-Ta-O bond angle distribution (ADF) in the grown oxide film for molecular/natural oxidation. (c) Ta-O RDF, and (d) O-Ta-O ADF for atomic oxidation. The distributions are shifted upwards for each temperature for better visualization.

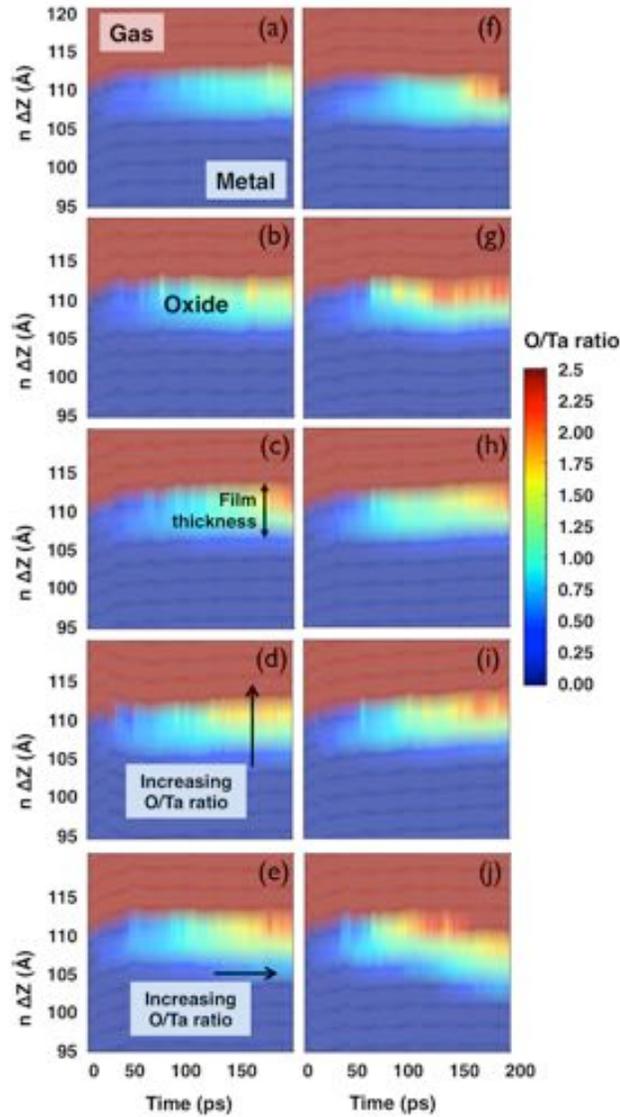

**Figure 8:** Evolution of O/Ta stoichiometry in the growing oxide film as a function of oxidation time for the oxidation of the bare Ta substrate at various temperatures. The 2-d color-maps have time (in ps) along the horizontal-axis, the spatial coordinate normal to the metal substrate surface along the vertical-axis, and is colored by the O/Ta stoichiometry ratio. Panels (a) to (e) are for molecular/natural oxidation with increasing temperature (300, 400, 500, 600, 900 K). Panels (f) to (j) are corresponding plots for atomic oxidation. The spatial bin size (vertical axis) is ~2.4 Å; the horizontal stripes are, thus, renderings of the bin boundaries. The gas region (identified by absence of Ta atoms) is colored maroon.

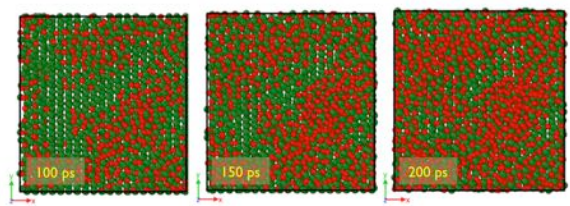

**Table of Contents Graphic**